\documentclass[aps,pra,reprint,groupedaddress,showpacs]{revtex4-1}
\usepackage{amsmath,amssymb,graphicx}
\usepackage{hyperref}
\def\id{{\mathchoice {\rm 1\mskip-4mu l} {\rm 1\mskip-4mu l} {\rm
1\mskip-4.5mu l} {\rm 1\mskip-5mu l}}}

\begin{document}

\title{Space-time qubits}

\author{J.L. Pienaar}
 \email{j.pienaar@physics.uq.edu.au}
\author{C.R. Myers}
 \email{myers@physics.uq.edu.au}
\author{T.C. Ralph}
 \email{ralph@physics.uq.edu.au}
\affiliation{
 Centre for Quantum Computation and Communication Technology, \\ 
 School of Mathematics and Physics, The University of Queensland, Brisbane 4072 QLD Australia
}

\date{\today}

\begin{abstract}
We construct a qubit algebra from field creation and annihilation operators acting on a global vacuum state. Particles to be used as qubits are created from the vacuum by a near-deterministic single particle source. Our formulation makes the space-time dependence of the qubits explicit, preparing the way for quantum computation within a field framework. The method can be generalized to deal with interacting qubits whose wavepackets are not perfectly matched to each other. We give an example of how to calculate the Heisenberg evolution of a simple two-qubit circuit, taking expectation values in the field vacuum state.
\end{abstract}


\maketitle

\section{Introduction}
Much of quantum information theory is couched in the formalism and language of the Schr\"odinger Picture of quantum mechanics, in which the qubit states explicitly evolve as they pass through circuits or down communication channels \cite{NIE}. Key properties such as entanglement depend specifically on the Hilbert space structure of the qubit state after processing. In contrast, a completely equivalent approach is to work in the quantum mechanical Heisenberg Picture, in which it is the Pauli operators describing observables that evolve, whilst the state of the system is  always the global ground-state of the qubits \cite{DEU00}. In this formalism the properties of the system depend on the Pauli operator algebra and specifically on its commutator structure. Heisenberg evolution is used extensively in the stabilizer formalism for error correction and graph states \cite{GOT98}.

A limitation of current quantum information theory is that it treats qubits as stationary point particles possessing no space-time extent or dynamics. A specific exception is in optics where the finite extent and propagation of photon modes has been included in linear optical quantum computation circuits \cite{2ROH05}.

In this paper we present a recipe for performing quantum information calculations in the Heisenberg Picture in which the Pauli operators are explicitly constructed from field operators with a spacetime structure and the initial state is always the global field ground, or vacuum state. This is in contrast to the usual approach, where the ground state is an n-qubit state, rather than the vacuum state. The new approach requires us to express the usual creation and annihilation operators, which act on the n-qubit ground state, as a function of the field operators, which act on the field vacuum state. To derive such an expression, we use an explicit model of a single-photon source \cite{MIG02} to treat the deterministic creation of a qubit from the vacuum.

The paper is arranged in the following way: in section II we show how a qubit algebra can be constructed from the vacuum using the creation and annihilation operators of a bosonic field with explicit space-time parameterization. In keeping with the Heisenberg Picture throughout, we find it necessary in section III to introduce an operator corresponding to the deterministic creation of a single particle mode. This allows us to represent arbitrary quantum gates acting on a collection of qubits as operators on the vacuum state. In section IV we show that the algebra is sufficiently general to cope with operations on qubits whose wavepackets are not perfectly matched to one another. Finally, in section V we illustrate these methods by applying them to the simple problem of a c-not gate in flat space-time. We conclude by discussing the potential applications of this work to problems in relativistic quantum information theory.

\section{Field Qubits}
We will consider a field of scalar bosons, which may be expanded in terms of its positive and negative frequency parts as:
\begin{equation}\label{bosonfield}
 \begin{aligned}
  \hat{\phi}(x,t) &= \int \mathrm{d}\textbf{k} \; [G(k,x)\hat{a}_{\textbf{k}}+G^*(k,x)\hat{a}^\dagger_{\textbf{k}}] \\
                       &\equiv \hat{A}_{G}+\hat{A}^\dagger_{G} \, ,
 \end{aligned}
\end{equation} 
where $G(k,x)$ is a normalized solution of the Klein-Gordon equation and $x,k$ are 4-vectors, with $kx = g_{\mu \nu}k^{\mu}x^{\nu}$. We will find it useful to consider wavepackets, which are localized superpositions of plane wave solutions:
\begin{equation}\label{packet}
G(k,x) = g(\textbf{k})e^{ikx} \, .
\end{equation}
where $g(\textbf{k})$ is centered on some positive wave number $k_0$ and is required to be zero for $k < 0$. Normalization requires:
\[  \int \mathrm{d}\textbf{k} \; |g(\textbf{k})|^2 = 1 \, . \]
We then interpret the operator $\hat{A}_{G}$ as the annihilation operator for a particle in the mode $G$, and $\hat{A}^\dagger_{G}$ as the corresponding creation operator. 
From the bosonic commutation relation $[\hat{a}(k),\hat{a}^\dagger(k')]=\delta(k-k')$ we obtain the same-time wavepacket commutator:
\begin{equation}\label{wavcom}
 \begin{aligned}
  &\left[ \hat{A}_{G}(k,\textbf{x}_1,t), \hat{A}^\dagger_{H}(k',\textbf{x}_2,t) \right] \\
   &= \int \mathrm{d}\textbf{k} \; G(k,\textbf{x}_1,t)H^*(k,\textbf{x}_2,t) \\
   & = \int \mathrm{d}\textbf{k} \; h^*(\textbf{k})g(\textbf{k})e^{\textbf{x}_1-\textbf{x}_2} \, . 
  \end{aligned}
 \end{equation}
In the rest of this section, we will restrict ourselves to a set of modes $G_i$ that are either perfectly matched in their spectral and spatial degrees of freedom, or perfectly orthogonal to each other, i.e. 
\[ [\hat{A}_{G_i},\hat{A}^\dagger_{G_j}] = \int \mathrm{d}\textbf{k} \; g_i(\textbf{k})g^*_j(\textbf{k}) = \delta_{ij} \, . \]
Hence the particular spatial dependence of the modes plays no role for the time being, so we will drop the subscript $G$ and merely consider the index $i$, hence $\hat{A}_{G_i} \rightarrow \hat{A}_{i}$. The spatial dependence will be re-introduced in section IV, where we show that a qubit algebra can still be constructed in a straightforward manner using information about the particles' spatial profiles; for the moment, we describe the construction of a simplified qubit algebra from ideal modes with trivial space-time dependence.
To consider qubits, it is convenient to work in the Fock basis and write a single particle state as the action of $\hat{A}^\dagger_{i}$ on the vacuum state:
\[ |1_i\rangle \equiv \hat{A}^\dagger_{i}|0\rangle \, . \]
To build up a qubit algebra, we need to start from a `qubit ground state' which is an $n$ particle state in the field representation. In the remainder of this section, we construct an algebra assuming that we have such an $n$ particle state. Section III will then illustrate how particle creation can be incorporated in the algebra.
Suppose that we have $n$ particles, each confined to a pair of orthogonal modes, $\hat{A}^\dagger_{i},\hat{B}^\dagger_{i}$, where $i=1,2,..,n$ labels the $i_{th}$ particle. Provided each particle remains confined to its two modes, we may interpret the particles as qubits in a dual-rail encoding and the state as an $n$-qubit system. For the $i_{th}$ qubit in the two-mode number basis $|n_A n_B\rangle_i$, we define the computational basis states $|\textbf{0}\rangle,|\textbf{1}\rangle$:
\[ |\textbf{0}\rangle_i \equiv \hat{A}^\dagger_i |00\rangle \equiv |10\rangle_i; \quad |\textbf{1}\rangle_i \equiv \hat{B}^\dagger_i |00\rangle \equiv |01\rangle_i \, . \]
An arbitrary superposition state is: 
\[ |\psi_i \rangle \equiv \alpha|\textbf{0}\rangle_i + \beta|\textbf{1}\rangle_i =\alpha |10\rangle_i + \beta |01\rangle_i ; \]
\[ (|\alpha|^2+|\beta|^2=1) \, . \]
We consider the following operations on the $i_{th}$ qubit:
\[ \hat{I}_{i}=\hat{A_i}^\dagger \hat{A_i}+\hat{B_i}^\dagger \hat{B_i}  \]
\[ \hat{Z}_{i}=\hat{A_i}^\dagger \hat{A_i}-\hat{B_i}^\dagger \hat{B_i}  \]
\[ \hat{X}_{i}=\hat{A_i}^\dagger \hat{B_i}+\hat{B_i}^\dagger \hat{A_i}  \]
\begin{equation} \label{pauli}
\hat{Y}_{i}=\textbf{\textit{i}}\hat{B_i}^\dagger \hat{A_i}-\textbf{\textit{i}}\hat{A_i}^\dagger \hat{B_i} \, .
\end{equation}
These are the quantum Stokes operators \cite{KOR02} . They preserve the qubit structure of the state space, and within it they form a representation of the familiar Pauli algebra when expectation values are taken. For example, even though the operator $\hat{X}\hat{X} = \hat{A}^\dagger \hat{B}\hat{A}^\dagger \hat{B}+\hat{B}^\dagger \hat{A}\hat{A}^\dagger \hat{B}+\hat{A}^\dagger \hat{B}\hat{B}^\dagger \hat{A}+\hat{B}^\dagger \hat{A}\hat{B}^\dagger \hat{A}$ acting on the $i_{th}$ qubit looks very different to $I=\hat{A}^\dagger \hat{A}+\hat{B}^\dagger \hat{B}$, a quick calculation shows that these operators are equivalent under expectation values taken in the state $|\psi_i \rangle$. It is therefore convenient to replace the operators in (\ref{pauli}) with their equivalence classes, which are formally representible by the Pauli matrices:
\[
 \hat{I}_{i}= \left[ \begin{array}{*{2}{c}}
    1 & 0 \\
    0 & 1
   \end{array} \right]_i 
\]
\[
 \hat{X}_{i}= \left[ \begin{array}{*{2}{c}}
    0 & 1 \\
    1 & 0
   \end{array} \right]_i 
\]
\[
 \hat{Y}_{i}= \left[ \begin{array}{*{2}{c}}
    0 & -\textbf{\textit{i}} \\
    \textbf{\textit{i}} & 0
   \end{array} \right]_i 
\]
\[
 \hat{Z}_{i}= \left[ \begin{array}{*{2}{c}}
    1 & 0 \\
    0 & -1
   \end{array} \right]_i 
 \, . \]
This allows us to perform calculations using just the Pauli operators, without having to refer to the fields themselves. At the end of the calculation, we need a recipe to convert the final matrix back into field operators in order to calculate the expectation value. Motivated by the correspondence between the field expressions (\ref{pauli}) and their matrix counterparts, we propose that a general single qubit matrix acting on the $i_{th}$ qubit can be decomposed into fields as:
\begin{equation}
    \left[ \begin{array}{*{2}{c}}
    \gamma & \delta \\
    \rho & \sigma
   \end{array} \right]_i \equiv \gamma \hat{A_i}^\dagger \hat{A_i} + \delta \hat{A_i}^\dagger \hat{B_i} + \rho \hat{B_i}^\dagger \hat{A_i} + \sigma \hat{B_i}^\dagger \hat{B_i} \, .
\end{equation}
Using the matrix tensor product, we may generalize this decomposition to $2^n \times 2^n$ matrices acting on the entire $n$-qubit system:
\begin{equation} \label{matrix}
 \begin{aligned}
  &\left[ \begin{array}{*{3}{c}}
    \gamma & \delta & \hdots \\
    \rho & \sigma & \hdots \\
    \vdots & \vdots & \kappa
   \end{array} \right]_i \\
  &\equiv \gamma \hat{A_1}^\dagger \hat{A_1}\hat{A_2}^\dagger \hat{A_2}...\hat{A_n}^\dagger \hat{A_n}\\ 
  &+ \delta \hat{A_1}^\dagger \hat{A_1}\hat{A_2}^\dagger \hat{A_2}...\hat{A_n}^\dagger \hat{B_n} \\
  &+...+\kappa \hat{B_1}^\dagger \hat{B_1}\hat{B_2}^\dagger \hat{B_2}...\hat{B_n}^\dagger \hat{B_n} \, .
 \end{aligned}
\end{equation}
The $n$-qubit identity in $n$-qubit space is therefore defined by:
\begin{equation} \label{identity}
 \hat{1} \equiv \hat{I}_{1} \otimes \hat{I}_{2} \otimes ... \otimes \hat{I}_{n} \, .
\end{equation}
In order to model general $n$-qubit gates we require a universal gate set. We can construct arbitrary single qubit operations from the Pauli operators by (for example) combining rotations around the Z axis, $\hat{Z}(\theta) = \cos{\theta} \hat I + i \sin{\theta} \hat{Z}$, with the Hadamard gate, $\hat{H} = 1/\sqrt{2}(\hat{Z} + \hat{X})$. We also need to add a two-qubit controlled sign (c-sign) gate to complete the set \cite{NIE}. 
In 2-qubit space, the c-sign has the matrix representation (in the computational basis):
\begin{equation}
 \begin{aligned}
  \bar{U}_{csign} &= \frac{1}{2}\left(\hat{I}_1\hat{I}_2+\hat{I}_1\hat{Z}_2+\hat{Z}_1\hat{I}_2-\hat{Z}_1\hat{Z}_2 \right) \\
  &= \left[ \begin{array}{*{4}{c}}
    1 & 0 & 0 & 0 \\
    0 & 1 & 0 & 0 \\
    0 & 0 & 1 & 0 \\
    0 & 0 & 0 & -1
   \end{array} \right] \, .
 \end{aligned}
\end{equation}
In terms of fields, using (\ref{matrix}) and (\ref{identity}), this matrix corresponds to the operator:
\begin{equation} \label{csign}
= \hat{1} - 2\hat{B_1}^\dagger \hat{B_1}\hat{B_2}^\dagger \hat{B_2} \, .
\end{equation}
Taking this as our c-sign gate completes the universal gate set. Arbitrary $n$-qubit gates can be represented as tensor products of the Pauli gates (\ref{pauli}) and c-sign gate (\ref{csign}) acting on one- and two-qubit subspaces. We can also translate between the familiar matrix representations and the field operator representations using (\ref{matrix}).

\subsection{Heisenberg Picture}
We now briefly review how quantum circuit calculations can be performed in the Heisenberg Picture \cite{DEU00,GOT98}. Instead of starting with some initial state, $|\psi \rangle$ that is evolved to some final state, $\hat U|\psi \rangle$ as in the Schr\"odinger Picture, we start from some set of observables representing the measurements made on the system, $\hat J$, and evolve them back to their initial values, $\hat U^{\dagger} \hat J \hat U$, where expectation values can be taken against the initial state, $\langle \psi| \hat U^{\dagger} \hat J \hat U |\psi \rangle$. The measurement operators are constructed from the Pauli operators, i.e. $\hat J = f(\hat I, \hat X, \hat Y, \hat Z)$ and, given that the evolution operators are also constructed from Pauli's, i.e. $\hat U = g(\hat I, \hat X, \hat Y, \hat Z)$, calculation of $\hat U^{\dagger} \hat J \hat U$ is straightforward from the Pauli commutation relations.

Clearly the two pictures are equivalent with respect to expectation values as $\langle \psi| (\hat U^{\dagger} \hat J \hat U) |\psi \rangle = (\langle \psi| \hat U^{\dagger}) \hat J (\hat U |\psi \rangle$). Here we are interested in ``pure" Heisenberg evolution for which all evolution, including state preparation, is performed in the Heisenberg Picture and so expectation values are always taken with respect to the qubit ground-state, i.e. $\langle \hat J \rangle = \langle 0| \hat U^{\dagger} \hat J \hat U |0 \rangle$.

\section{Single particle production}
\begin{figure}[!htbp]\begin{center}
\includegraphics[width=8cm]{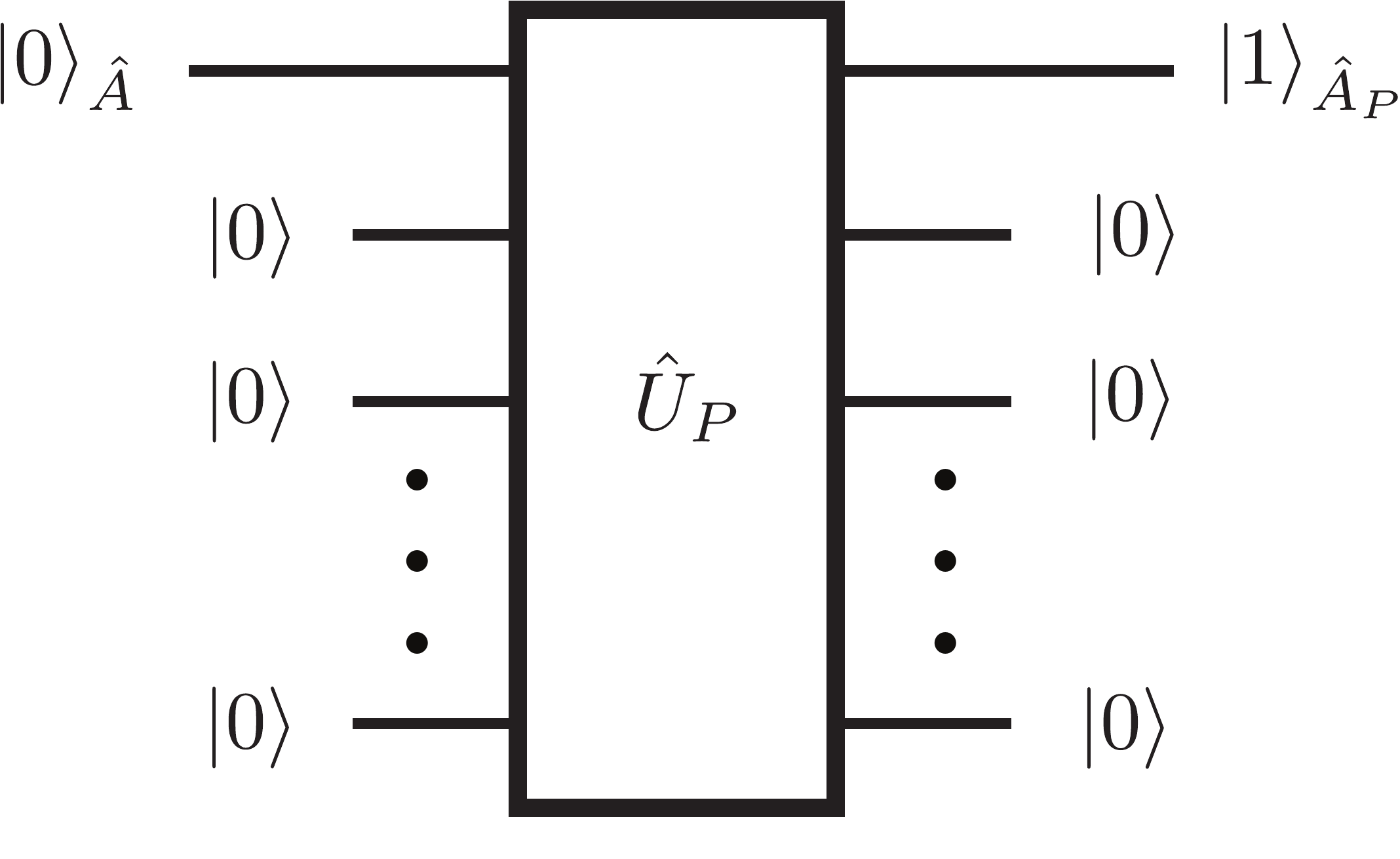}
\caption{Schematic of a unitary single particle source. We will find it useful to make use of a large number of ancilla states, which begin and end in the vacuum state.}
\label{Usp}
\end{center}\end{figure}

The qubit ground-state in the preceding section is an $n$-particle state (where $n$ is the number of qubits) and so is not the field ground-state. In order to have ``pure" Heisenberg evolution of the field state we need to identify the Heisenberg transformation representing single particle creation. Strictly this means finding a unitary transformation that takes us from the vacuum to the $n$-qubit state. Ideally, we would like to have a unitary $\hat{U}_P$ that takes a vacuum mode $\hat{A}$ as input and gives a single particle in the mode $\hat{A}_P$ as output. This is shown schematically in Fig. \ref{Usp}. 

Since the input and the output are both pure states, such a unitary must exist in principle. We could then obtain the output mode $\hat{A}_P$ as a function of the input vacuum modes by calculating its Heisenberg evolution:
\begin{equation}
 \hat{U}^\dag \hat{A} \hat{U} \equiv \hat{A}_P \, .
\end{equation}

\begin{figure}[!htbp]\begin{center}
\includegraphics[width=8cm]{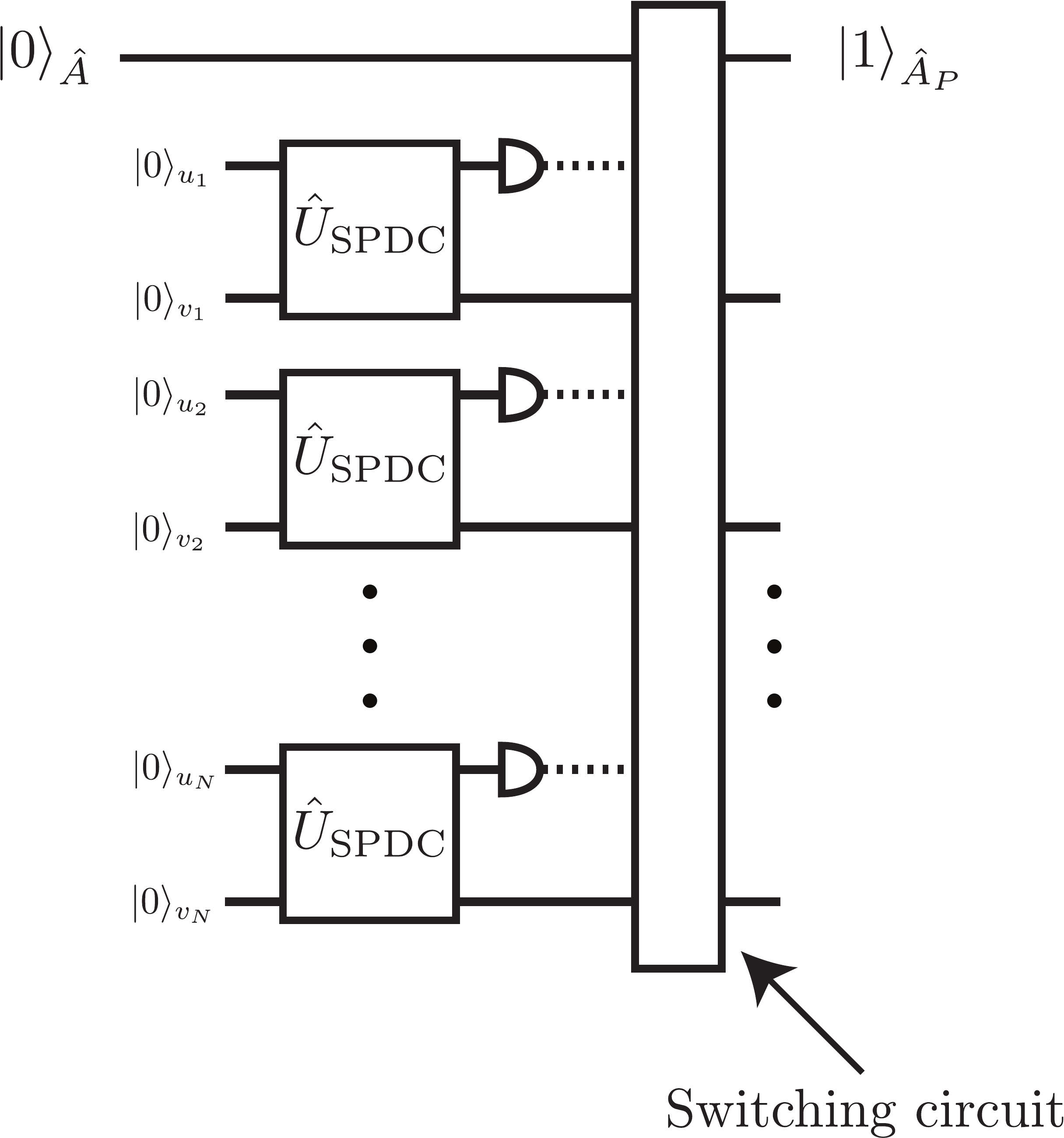}
\caption{A non-unitary near-deterministic single particle source.}
\label{mig}
\end{center}\end{figure}

However, finding an explicit form for the unitary is a nontrivial matter because it is a highly non-linear function of the field operators. Rather than attempt such a task, we will consider a physical model that produces the exact same evolution using detection and feed-forwards. The model will give us an explicit recipe for writing down the output mode $\hat{A}_P$ in terms of vacuum input modes. Specifically, we refer to a suggestion by Migdall, Branning and Castelletto \cite{MIG02} for a deterministic single photon source that makes use of $N$ spontaneous parametric down-converters (SPDCs) and post-selection (see Fig. \ref{mig}) to produce the desired evolution in the limit of arbitrarily large $N$. The basic model is not specific to optics and we will present a slightly more abstract  version that is well suited to our purposes.

\subsection{The Migdall-Branning-Castelletto Source}\label{MigdallSourceSection}
Consider the parametric amplification unitary:
\[ \hat{U}_{P}\equiv exp \iint\mathrm{dk}\mathrm{dk '} [\chi P(k,k ') \hat{u}^\dagger(k)\hat{v}^\dagger(k ') - H.c.] \]
This unitary acts on the vacuum to create a two particle state with probability $\chi^2$; one particle in the mode $\hat{u}$ and one in the mode $\hat{v}$ (the $\hat{u}$ particle may be detected to herald the presence of the $\hat{v}$ particle). $\chi$ is assumed $\ll 1$ and so terms of higher order than $\chi^2$ are neglected. The particles have a joint spectral amplitude denoted $P(k,k ')$ which in general includes entanglement between the particle pairs; however for simplicity we will assume that the state is separable (achieved in the lab by phase-matching) and has the form $P(k,k')=G(k)G(k')$ \cite{GRI97}. Under this assumption there are no multimode effects, so we can use the shorthand
\[ \int \mathrm{dk} G(k) e^{i(kx-\omega t)}\hat{u}(k) \rightarrow \hat{u} \, , \]
and similarly for the mode $\hat{v}$. The Heisenberg evolution of the operators is then:
\begin{equation}\label{Eqn:HeisenbergDC}
 \begin{aligned}
  \hat{u}' = \hat{U}^\dagger_{P}\hat{u}\hat{U}_{P} = sinh(\chi)\hat{u}^{\dagger}+cosh(\chi)\hat{v}\\
  \hat{v}' = \hat{U}^\dagger_{P}\hat{v}\hat{U}_{P} = sinh(\chi)\hat{v}^{\dagger}+cosh(\chi)\hat{u} \, .
 \end{aligned} 
\end{equation}
Generalizing to an array of $N$ such unitaries, labelled from 1 to $N$, we take the $i_{th}$ unitary to act on the modes $\hat{u_i}$,$\hat{v_i}$ with an output amplitude $G^*_i(k)G^*_i(k')$. The modes are assumed to be mutually orthogonal and satisfy the commutation relations:
\[ [\hat{u}_i,\hat{u}^\dagger_j] = \delta_{ij} , \quad [\hat{v}_i,\hat{v}^\dagger_j] = \delta_{ij} \, . \]
The output of the array is:
\begin{equation} \label{pdcState}
  |\psi_P \rangle \equiv \hat{U}_{P_1}\otimes\hat{U}_{P_2}\otimes...\hat{U}_{P_N}|0\rangle \, .\\ 
\end{equation}
Although the probability of any individual SPDC firing remains of order $\chi^2$, where $\chi^2$ is small, the probability of at least one of the SPDCs in the array firing scales with $N$, so that for $N$ sufficiently large, we are almost certain to have at least one particle available at any given moment. Furthermore, $g^{(2)}$ does not scale with $N$ because it represents the relative probability of obtaining more than one photon for a given mode, independently of modes other than the one being considered. Hence the probability of obtaining two or more particles in the \textit{same} mode remains negligible. 
Finally, we perform detection on the $\hat{u}_i$ modes in (\ref{pdcState}) and feed forward the topmost heralded particle in the array into the mode $\hat{A}_P$. All other particles that may have been produced are dumped and replaced with empty modes. The information about the detection result for the mode $\hat{u}_i$ is contained in the particle-number operator:
\[ \hat{n}_{u'_j} \equiv \hat{u}'^\dag_{j}\hat{u}'_{j} \]
and has eigenvalues 0,1 or 2. Higher values than 2 correspond to higher orders of $\chi$ and can be neglected provided we choose $\chi$ to be small, as we do in this paper. In terms of the feed-forward and the vacuum modes, we find that the output mode $\hat{A}_P$ has the form: 
\begin{equation} \label{ap}
  \hat{A}_P \equiv \sum_{j=1}^{N} \hat{d}_{u'_j} \hat{v}'_j \prod_{i=0}^{j-1} \left( 1 - \hat{d}_{u'_i} \right) + \hat{c}\hat{A} \, ,
\end{equation}
where $\hat{A}$ is the vacuum input mode referred to at the beginning of the section.
The expression depends explicitly on the feed-forward $\hat{n}_{u'_j}$ through a bucket-detector operator $\hat{d}_{u'_j}$:
\begin{equation}\label{ffwd} 
\hat{d}_{u'_j} \equiv \frac{1}{2}(3-\hat{n}_{u'_j})\hat{n}_{u'_j} \, .
\end{equation} 
Given the non-negligible eigenvalues of $\hat{n}_{u'_j}$, the operator (\ref{ffwd}) represents a detector that registers a single click when one or more particles are present and does nothing otherwise. This detector is unable to resolve particle numbers greater than one; this is in keeping with the bucket detector model.

\begin{figure}[!htbp]\begin{center}
\includegraphics[width=8.6cm]{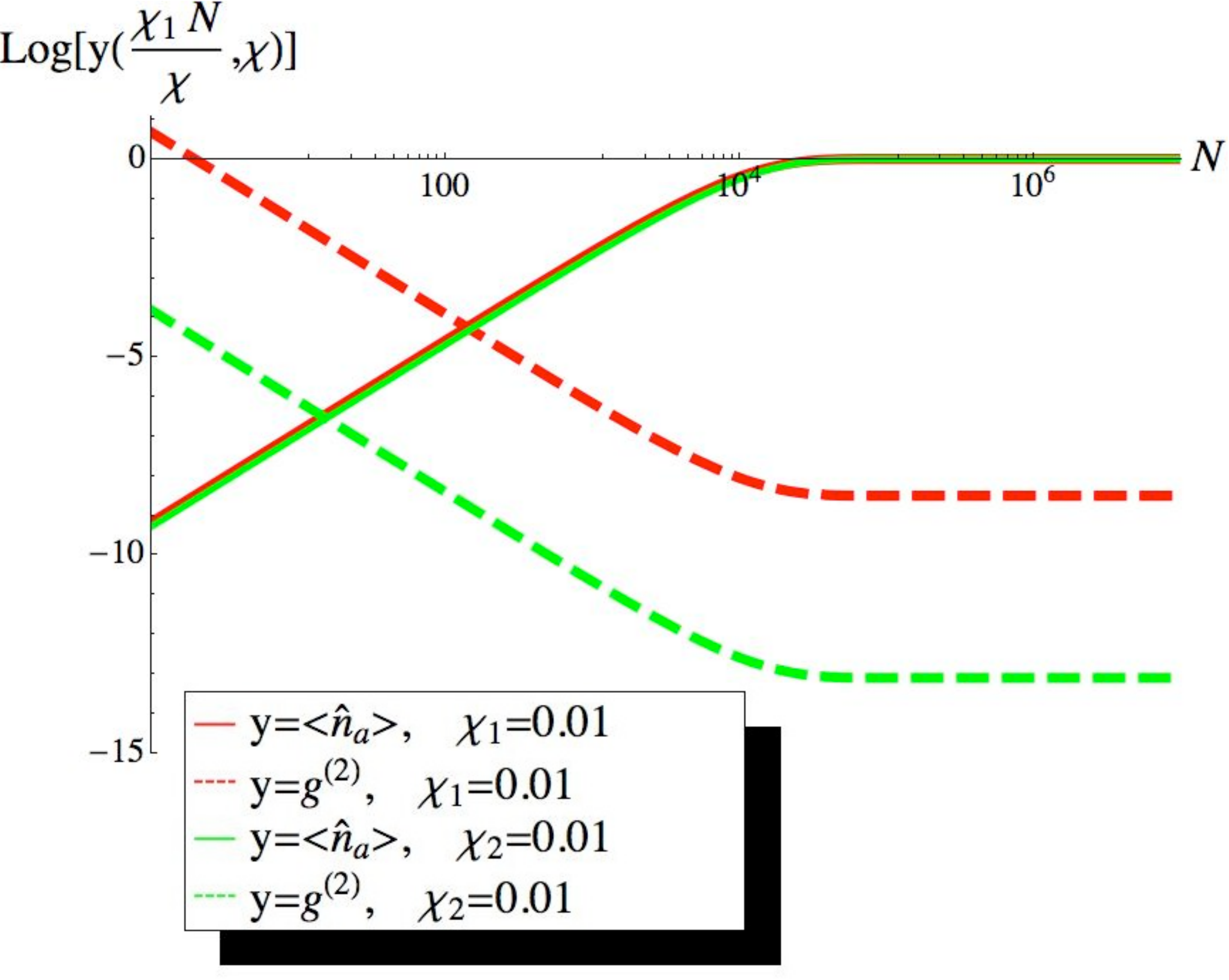}
\caption{(Colour online) Graph of $n_{A_P}$ and $g^{(2)}$ for $\chi_{1}=0.01$ (red) and $\chi_{2}=0.001$ (green) as a function of $N$. The horizontal scaling has been adjusted relative to $\chi_1=0.01$ to bring the graphs into coincidence for ease of viewing. The vertical scaling is the same for all graphs to allow for a comparison. While the graphs of $n_{A_P}$ always saturate at $n_{A_P}\simeq1$ for some $N$, the graphs of $g^{(2)}$ saturate at values that become closer to zero (minus infinity on the Log scale) as we make $\chi$ smaller. Thus we can choose $\chi$ so as to make $g^{(2)}$ as small as we like, while still obtaining $n_{A_P}\simeq 1$ for sufficiently large $N$.}
\label{graphs}
\end{center}\end{figure}

Any single particle mode must have a particle-number expectation value of 1, a negligible probability of containing more than one photon, and it should satisfy the usual commutation relation. Formally, these properties are given by the relations:
\begin{equation} \label{identities}
 \begin{aligned}
  & \langle 0|\hat{A}^\dag_{P}\hat{A}_{P}|0\rangle = 1, \\
  & \langle 0|\hat{A}^\dag_{P}\hat{A}^\dag_{P}\hat{A}_{P}\hat{A}_{P}|0\rangle=0, \\
  & [\hat{A}_{P},\hat{A}^\dag_{P}] = \delta(k-k ') \, .
 \end{aligned}
\end{equation}
For $\hat{A}_P$ to satisfy the commutation relation in (\ref{identities}), the operator $\hat{c}$ is taken to be:
\[ \hat{c} \equiv \left( \hat{1} - \sum_{j=1}^{N} \hat{n}^2_{u_j}\prod_{i=0}^{j-1} \left( 1 - \hat{n}_{u_i} \right)^2 \right)^{\frac{1}{2}} \, . \]
We need to verify that $\hat{A}_P$ also satisfies the other properties in (\ref{identities}). From a direct calculation, the details for which are shown in the Appendix, we find: 
\begin{equation} \label{nAp}
 \begin{aligned}
  \langle\hat{n}_{A_{P}}\rangle & =  \langle 0|\hat{A}^\dag_{P}\hat{A}_{P}|0\rangle \\
                                & =  \frac{(4-4|\chi|^2+9|\chi|^4)((1-|\chi|^2)^N-1)}{5|\chi|^4-4} \, ,
 \end{aligned}
\end{equation}
\begin{equation} \label{g2}
 \begin{aligned}
  g^{(2)} & =  \frac{ \langle 0|\hat{A}^\dag_{P}\hat{A}^\dag_{P}\hat{A}_{P}\hat{A}_{P}|0\rangle}{\langle\hat{n}_{A_{P}}\rangle^2} \\
                                & =  \frac{2|\chi|^2(4-5|\chi|^4)^2}{(4-4|\chi|^2+9|\chi|^4)^2((1-|\chi|^2)^N-1)} \, .
 \end{aligned}
\end{equation}
In the limit of $N$ large and $\chi$ small, one finds that these quantities approach 1 and 0 respectively (see Fig.\ref{graphs}); hence the properties (\ref{identities}) are indeed satisfied. In idealized situations, the full expression (\ref{ap}) is not needed; the identities (\ref{identities}) suffice for performing calculations. In more practical situations where particle production might be non-ideal, or when we may wish to consider relativistic observers (see Section VI), it will be necessary to have the explicit expression (\ref{ap}) for $\hat{A}_P$ in terms of the initial vacuum modes.

Explicitly, if we wish to enact a circuit $\hat{U}_C$ on $n$ qubits produced by the above method, where the circuit acts on the modes $\hat{A}_{G_1},\hat{A}_{G_2},...,\hat{A}_{G_n},\hat{B}_{G_1},\hat{B}_{G_2},...,\hat{B}_{G_n}$, then instead of calculating the usual general expectation value:
\begin{equation} \label{schro}
 \langle \psi_{in} |F(\hat{A}_{G_1},\hat{A}_{G_2},...,\hat{B}_{G_1},...)| \psi_{in} \rangle \, ,
\end{equation}
where $| \psi_{in} \rangle$ is the initial n-particle prepared state
\[ |\psi_{in}\rangle \equiv \hat{A}^\dagger_{G_1}\hat{A}^\dagger_{G_2}...\hat{A}^\dagger_{G_n}|0\rangle \, , \]
 we now calculate the expectation value:
\begin{equation}\label{heis}
 \begin{aligned}
  &\langle 0|F(\hat{A}_{P,G_1},\hat{A}_{P,G_2},...,\hat{B}_{P,G_1},...)|0\rangle \\
  = &\langle 0|f(\hat{u}_{G_1},\hat{u}_{G_2},...,\hat{v}_{G_1},...)|0\rangle \, ,
 \end{aligned}
\end{equation}
where the $\hat{A}_{P,G_i},\hat{B}_{P,G_i}$ are defined in terms of the vacuum modes $\hat{u}_{G_i},\hat{v}_{G_i}$ according to (\ref{ap}).

\section{Particle Mismatch}
We would now like to relax the assumption that particles in non-orthogonal modes have exactly the same mode distribution $G(k,x)$. In the context of wavepackets, this will introduce nontrivial commutators of the form (\ref{wavcom}). This more general situation might include interactions between particles whose spectral amplitudes are not the same, as in quantum optics, or situations in which the particles arrive at the gate at slightly different times, or their wavepackets only partially overlap in space. In the discussion that follows, we will focus on the latter case of mismatch caused by mis-timing or spatial discrepancies, but our analysis applies equally well to other types of mismatch.

One might worry that the algebra of section II no longer applies in this scenario, since the extra degrees of freedom might allow qubits to become `invisible' with respect to each other, making it unclear how to define an n-qubit space. We will make use of a proposal, made in the context of optical quantum computation, by Rohde, Mauerer and Silberhorn (RMS)\cite{ROH07}, which allows us to separate the amplitudes $G(k,x)$ into parts which are separately matched or orthogonal to the interaction of interest. We will show that this allows us to retain the Pauli algebra in general situations, at the cost of expanding the state space to include the orthogonal degrees of freedom.  

\begin{figure}[!htbp]\begin{center}
\includegraphics[width=8.6cm]{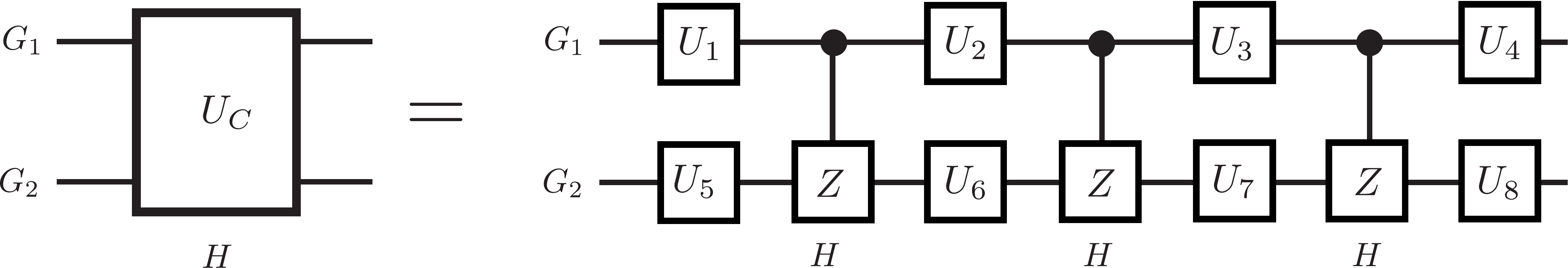}
\caption{A general unmatched two-qubit circuit. Here $G_{1,2}$ denotes $G(k,x_{1,2})$ and $H$ denotes $H(k,x_c)$.}
\label{circa}
\end{center}\end{figure}

Following RMS, the mismatch between particle modes is treated as a rotation into the orthogonal degrees of freedom with respect to some `reference' mode, with the amount of rotation being equivalent to the overlap between the amplitudes of the modes of interest with the reference mode. In this way we can write down an effective circuit that can be treated with all the usual  methods, yet with sufficient machinery to deal with potential qubit mismatch. We will demonstrate this technique for a general two-qubit gate.
The circuit in Fig.\ref{circa} is a two-qubit system, constructed from two particles with spectral amplitudes $G(k,x_1)$ and $G(k,x_2)$ and distributed amongst the pairs of modes $\hat{a}_{i,\vec{k}},\hat{b}_{i,\vec{k}}$, with $i=1,2$ labeling the qubits. Hence: $\hat{A}_{1,G_1}\equiv \int \mathrm{d}\vec{k} \; G(k,x_1) \;\hat{a}_{1,\vec{k}}$, etc. The qubits interact via a general two-qubit unitary $U_C$ whose Hamiltonian is assumed to depend on just a single amplitude $H(k,x_c)$ representing, for example, an external classical pump driving the interaction. Larger circuits containing interactions with different pump amplitudes can be constructed by connecting basic circuits of the type considered here, although each part would have to be treated separately. 

\begin{figure}[!htbp]\begin{center}
\includegraphics[width=8.6cm]{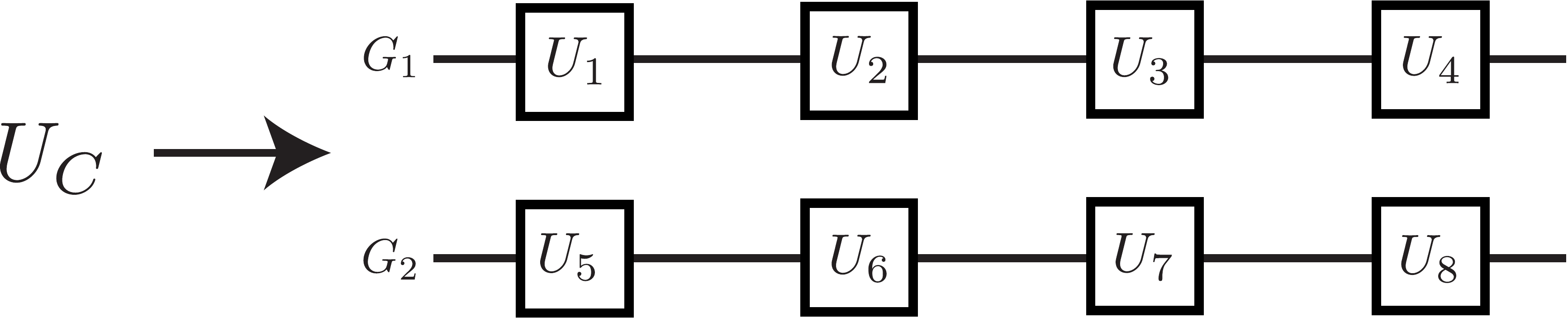}
\caption{The circuit as seen by orthogonal qubits.}
\label{circb}
\end{center}\end{figure}

Our task is to calculate the Heisenberg evolution through the circuit given that $G(k,x_1),G(k,x_2),H(k,x_c)$ are all different in general. We proceed by choosing the pump amplitude $H(k,x_c)$ as a reference, and describing all other modes by their overlap with $H(k,x_c)$. To this end, we will make use of a `mismatch factor' $\xi$, which we interpret as being the amount of overlap between the interacting wavepackets. Specifically, $\xi$ is determined by the commutator (\ref{wavcom}) between the appropriate wavepacket creation and annihilation operators at the time of interaction:
\begin{equation}\label{overlapterm}
 \begin{aligned}
  \xi \equiv |\int \mathrm{d}\textbf{k} \; G(k,\textbf{x}_1,t)H^*(k,\textbf{x}_c,t)|^2  \, .
 \end{aligned}
\end{equation}

It turns out that general situations in which neither qubit is matched to the reference amplitude can always be re-interpeted such that all of the mismatch is carried by only one of the qubits. For example, if the overlap terms are $\xi_1 = |\int\mathrm{d}\textbf{k} \, G(k,x_1)H^*(k,x_c)|^2$ and $\xi_2 = |\int\mathrm{d}\textbf{k} \, G(k,x_2)H^*(k,x_c)|^2$, we can multiply them to form a third quantity $\xi_3 \equiv \xi_1 \xi_2$ and then interpret $\xi_3$ as the mismatch of just one qubit whose partner is perfectly matched to the reference amplitude. Hence we lose no generality by assuming that one of the qubits is perfectly matched and the other is not. We will therefore let $G(k,x_2)=H(k,x_c)$ and let the entire mismatch be carried by the overlap of $G(k,x_1)$ with $H(k,x_c)$. We first write the modes as superpositions of matched and orthogonal components:
\begin{equation} 
 \begin{aligned}
  &\hat{A}_{1,G_1} \equiv \sqrt{\xi}\hat{A}_{1,H}+\sqrt{1-\xi}\hat{c} \, ,\\ 
  &\hat{B}_{1,G_1} \equiv \sqrt{\xi}\hat{B}_{1,H}+\sqrt{1-\xi}\hat{d},
 \end{aligned}
\end{equation}
where we have introduced the auxiliary vacuum modes $\hat{c},\hat{d}$. From a circuit perspective, we may view the $G_1$ modes as the result of a rotation that causes mixing between the matched and orthogonal parts. Formally, this is achieved by the evolutions:
\begin{equation}\label{subs}
 \begin{aligned}
  \hat{a}_1 &\rightarrow \sqrt{\xi}\, \hat{a}_1+\sqrt{1-\xi}\, \hat{c},\\
  \hat{c} &\rightarrow \sqrt{\xi}\, \hat{c}-\sqrt{1-\xi}\, \hat{a}_1,\\
  \hat{b}_1 &\rightarrow \sqrt{\xi}\, \hat{b}_1+\sqrt{1-\xi}\, \hat{d},\\
  \hat{d} &\rightarrow \sqrt{\xi}\, \hat{d}-\sqrt{1-\xi}\, \hat{b}_1 \, ,
 \end{aligned}
\end{equation}
where it is understood that the modes $\hat{a}_i,\hat{b}_i$ are matched to $H(k,x)$, i.e. we identify $\hat{a}_i \equiv \hat{A}_{i,H}(k,x)$, etc. The choice of phase convention in the above evolution doesn't matter because the modes representing the unmatched part are initially empty. We note that even if the spectral profiles of the interacting modes are identical, $h(\textbf{k})=g(\textbf{k})$, spatio-temporal differences between the modes such as that produced by a delay on one qubit may still give rise to a non-trivial overlap, since $G(k,x_1)\neq G(k,x_2)$ in general. We recover the algebra of section II as a special case: when $\xi=1$, then qubit 1 is perfectly matched and the orthogonal subspace is trivial; conversely if $\xi=0$ then qubit 1 is completely orthogonal and the qubit subspaces decouple.

The gate $U_C$ only acts on the matched modes $\hat{a}_1,\hat{b}_1,\hat{a}_2,\hat{b}_2$. The orthogonal modes $\hat{c},\hat{d}$ encounter a different circuit, $U'_C$, that can be obtained from $U_C$ by physical considerations, as we now show. To this end, we make use of the fact that any two-qubit unitary can be decomposed into single qubit gates and c-signs. Specifically, the general two-qubit unitary $U_C$ can be decomposed into eight single-qubit operations and three c-signs\cite{VID04}, as shown in Fig.\ref{circa}. The single-qubit gates may be constructed from (\ref{pauli}) as normal, using field modes matched to $G(k,x_1)$ and $H(k,x_c)$ for qubits 1 and 2 respectively. As the single-qubit operations are independent of one another, it makes no difference whether the qubits are matched or not; commutators between unmatched modes simply do not arise. The two-qubit c-sign gates are more complicated; we must make reference to a particular physical model. A natural choice, applicable to optical, microwave and ionic qubits\cite{MIL89,ROO08}, is the strong nonlinear cross-Kerr effect between two modes. To the external pump mediating this effect we prescribe the amplitude $H(k)$. The unitary operator for the c-sign acting on the modes $\hat{b}_1, \hat{b}_2$ is: 
\begin{equation} \label{kerr}
 \begin{aligned}
   \hat{U}_{Kerr} &= exp [ -i \pi \hat{n}_{b_1}\hat{n}_{b_2}] \\
                  &= \hat{1} + [ e^{-i \pi} - 1]\hat{n}_{b_1}\hat{n}_{b_2}\\
                  &= \hat{1} -2\hat{n}_{b_1}\hat{n}_{b_2}\\
                  &= \hat{1} -2\hat{B}^\dag_{H,1}\hat{B}_{H,1}\hat{B}^\dag_{H,2}\hat{B}_{H,2} \, ,
 \end{aligned}
\end{equation}
where we have used the property:
\[ \left( \hat{n}_{b_1}\hat{n}_{b_2} \right)^p \equiv \hat{n}_{b_1}\hat{n}_{b_2} \, , \]
which follows from that fact that the modes $\hat{b}_1, \hat{b}_2$ can each only contain 0 or 1 particle. Note that the unitary (\ref{kerr}) is consistent with an abstract c-sign (\ref{csign}) whose modes are matched to $H(k,x_c)$.
From  (\ref{csign}) we see that the absence of one or both particles implies $\hat{U}_{csign} \rightarrow \hat{1}$. This implies that the circuit $U'_C$ seen by the orthogonal modes $\hat{c},\hat{d}$ is just the decomposition of $U_C$ without the c-signs, see Fig.\ref{circb}. We can now interpret the evolution (\ref{subs}) as the action of a gate $U_{r}$ on the joint system $U_{C} \otimes U'_{C}$, which will be useful as we now show by example.

\section{An example: the C-NOT}
\begin{figure}[!htbp]\begin{center}
\includegraphics[width=8.6cm]{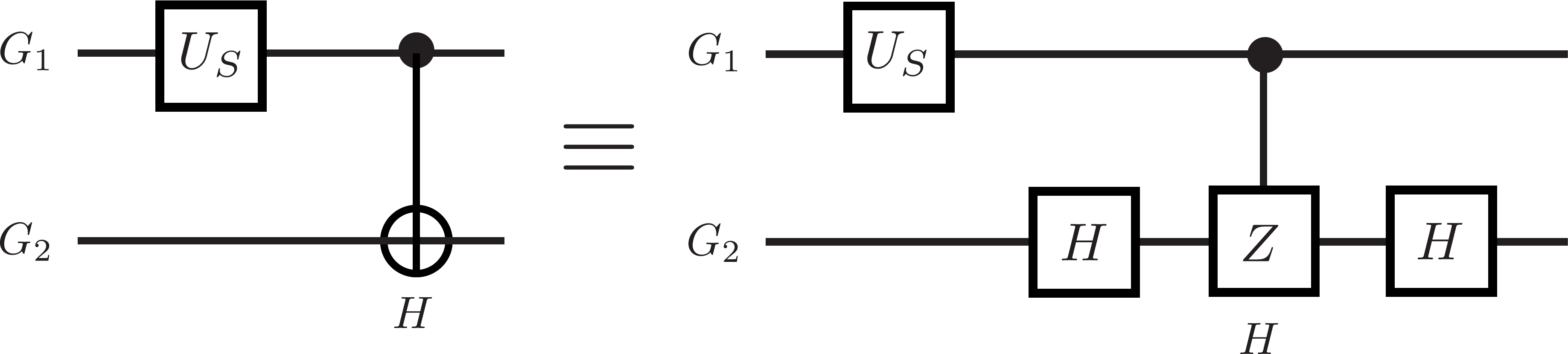}
\caption{A c-not between unmatched qubits, with an arbitrary rotation on qubit 1. The c-not gate is decomposed as a c-sign and two Hadamard gates.}
\label{cnota}
\end{center}\end{figure}

To illustrate the use of the techniques described above, we will calculate the output of a simple two-qubit circuit in which the qubit pulses have identical spectral properties but do not fully overlap at the gate. This could be achieved by placing a slight delay on one of the pulses, leading to a spatial discrepancy between the propagated modes at the time of interaction. We consider a circuit consisting of a single controlled NOT (c-not) gate, see Fig.\ref{cnota}. The first qubit is prepared in an arbitrary superposition by applying the operator $\hat{U}_s \equiv \sqrt{1-\alpha^2} \hat{I}_1 - \alpha \textbf{\textit{i}}\hat{Y}_1$. The output of this circuit in the Schr\"odinger picture for the ideal matched case is the Bell state: $\sqrt{1-\alpha^{2}}|00\rangle+\alpha|11\rangle$. We will now investigate how the spatial mismatch affects the functioning of the gate.

\begin{figure}[!htbp]\begin{center}
\includegraphics[width=8.6cm]{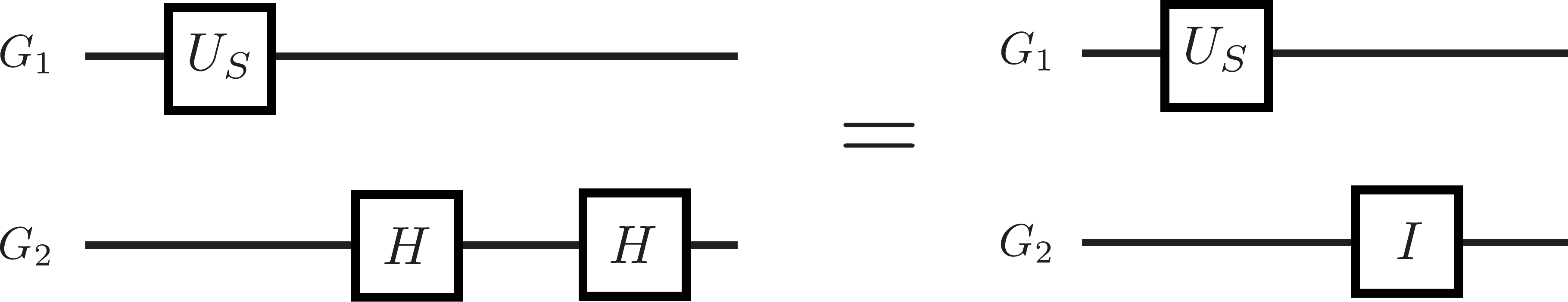}
\caption{The circuit as seen by orthogonal qubits.}
\label{cnotb}
\end{center}\end{figure}

For this example we will assume that the qubits are spectrally matched to each other and the gate; $g(\vec{k})=h(\vec{k})$. The mismatch is introduced by defining $G(k,x_1)=g(\vec{k})e^{(i\vec{k} \vec{x}_1-\omega t)}$ for qubit 1 and $G(k,x_2)=g(\vec{k})e^{(i\vec{k} \vec{x}_2-\omega t)}$ for qubit 2 and the gate interaction, where the modes are now the same apart from being centered at different points in space, $x_1 \neq x_2$, at time $t$.
From (\ref{overlapterm}), we find that the mismatch $\xi$ is:
\begin{equation}\label{overlap2}
  \xi \equiv | \int \mathrm{d}\textbf{k} \; g(\vec{k})g^*(\vec{k}) e^{i \vec{k}(\textbf{x}_1-\textbf{x}_2)}|^2 \, .
\end{equation}
The degrees of freedom are denoted by the orthogonal modes:
\begin{equation}\label{modes}
 \{ \hat{a}_1,\hat{b}_1,\hat{c},\hat{d},\hat{a}_2,\hat{b}_2 \} \, .
\end{equation}
The complete circuit containing $U_{cnot}$ is shown in Fig.\ref{cnota}, whereas the circuit seen by the orthogonal modes is shown in Fig.\ref{cnotb}. The c-not gate in the latter circuit, which we denote $U'_{cnot}$, is found to be the two-qubit identity using the methods of the preceding section. The modes $\hat{a}_1,\hat{a}_2$ are single particle modes and have the form (\ref{ap}) because the initial state is $|\textbf{00}\rangle$, which corresponds to the creation of a particle in each of the modes $\hat{a}_1$ and $\hat{a}_2$. 
The detectors are taken to be bucket detectors. Hence a measurement of the expectation value of the Pauli operator $\hat{X}$ on qubit 1 is represented by calculating the expectation value of the operator $\hat{X}^{(ab)}_{1}+\hat{X}^{(cd)}_{1}$ where 
\[ \hat{X}^{(cd)}_{1} \equiv \hat{c}^\dagger \hat{d}+\hat{d}^\dagger \hat{c},\quad \textrm{etc.} \]
All unitary operators acting on the qubit 1 subspace can be similarly decomposed into a sum of matched and orthogonal parts, corresponding to the $\hat{a}_1,\hat{b}_1$ and $\hat{c},\hat{d}$ subspaces respectively. Given the algebra (\ref{pauli}) together with knowledge of the orthogonal circuit Fig.\ref{cnotb}, we can construct a matrix representation of general two-qubit operations in the space spanned by $\{ \hat{a}_1,\hat{b}_1,\hat{c},\hat{d}\} \otimes \{ \hat{a}_2,\hat{b}_2 \}$ using $8\times 8$ matrices in block diagonal form in which the two blocks represent the matched and orthogonal subspaces: 
\[ \bar{U}_{C} = \left[ \begin{array}{*{2}{c}}
    \hat{U}_{C} & 0 \\
    0 & \hat{U}'_{C} 
   \end{array} \right] \, .
\]
To avoid confusion, we will denote operators acting on the entire space with a bar on top, $\bar{J}_1$, instead of a hat, $\hat{J}_1$, reserving the latter for operators acting on either subspace (since qubit 2 is perfectly matched, the unmatched subspace is trivial for that qubit: $\bar{K}_2 = \hat{K}_2$). We now write down the matrices for the gates in the circuit (\ref{cnota}):
\[ \bar{U}_{cnot} = \left[ \begin{array}{*{2}{c}}
    \hat{U}_{cnot} & 0 \\
    0 & \hat{I}_1\hat{I}_2
   \end{array} \right] \, ,
\]
\[ \bar{U}_{s} = \left[ \begin{array}{*{2}{c}}
    \hat{U}_{s} \hat{I}_2 & 0 \\
    0 & \hat{U}_{s} \hat{I}_2
   \end{array} \right] \, ,
\]
\[ \quad  (\text{where} \quad \hat{U}_{s} \equiv \left[ \begin{array}{*{2}{c}}
    \sqrt{1-\alpha^2} & -\alpha \\
    \alpha & \sqrt{1-\alpha^2}
   \end{array} \right] ) \, ,
\]
\[ \bar{U}_{r} = \left[ \begin{array}{*{2}{c}}
    \sqrt{\xi} \; \hat{I}_1\hat{I}_2 & \sqrt{1-\xi} \; \hat{I}_1\hat{I}_2 \\
    -\sqrt{1-\xi} \; \hat{I}_1\hat{I}_2 & \sqrt{\xi} \; \hat{I}_1\hat{I}_2
   \end{array} \right] \, .
\]
Intuitively, the matrix forms of $\bar{U}_{cnot}$ and $\bar{U}_{s}$ can be understood in terms of their action on the matched and orthogonal subspaces; the c-not acts differently on each subspace due to the varying behavior of the non-linear C-SIGN gate, but the rotation is just a linear single-qubit gate and works regardless of whether the qubit has been rotated into orthogonal modes or not. The unitary $\hat{U}_{r}$ is simply the rotation of the $\hat{a}_1,\hat{b}_1$ modes into the $\hat{c},\hat{d}$ modes according to (\ref{subs}). We can now calculate the expectation value of some arbitrary operator $\bar{J}$ on qubit 1 and $\bar{K}$ on qubit two. This operator may be represented by the matrix:
\[ \bar{J}_1\bar{K}_2 = \left[ \begin{array}{*{2}{c}}
    \hat{J}_1\hat{K}_2 & 0 \\
    0 & \hat{J}_1\hat{K}_2
   \end{array} \right]
\]
We wish to calculate the quantity
\begin{equation} \label{expec}
 \langle 0|\bar{U}^\dag_{sys} \bar{J}_1 \bar{K}_2 \bar{U}_{sys} |0 \rangle \, ,
\end{equation}
where $\bar{U}_{sys}$ is the combined evolution of the circuit,
\begin{equation}
 \begin{aligned}
  \bar{U}_{sys} \equiv \bar{U}_{cnot}\bar{U}_{bs}\bar{U}_{r} \, .
 \end{aligned}
\end{equation}
We find:
\[ \bar{U}_{sys} = \left[ \begin{array}{*{2}{c}}
    \sqrt{\xi} \hat{U}_{cnot} \hat{U}_{1bs} \hat{I}_2 & \sqrt{1-\xi} \hat{U}_{cnot} \hat{U}_{1bs} \hat{I}_2 \\
    -\sqrt{1-\xi} \hat{U}_{1bs} \hat{I}_2 & \sqrt{\xi} \hat{U}_{1bs} \hat{I}_2
   \end{array} \right] \, .
\]
For a given  $\bar{J}$ and $\bar{K}$, we can then obtain the matrix: 
\begin{equation} \label{mat}
\bar{U}^\dag_{sys} \bar{J}_1 \bar{K}_2 \bar{U}_{sys} \, .
\end{equation}
Taking the expectation value of this operator in the vacuum is trivial since the only modes containing particles are the $\hat{a}_1$ and $\hat{a}_2$ modes, so all matrix elements containing other terms annihilate the vacuum and vanish. The only remaining term is the coefficient of 
\[ \langle 0|\hat{a}^\dag_1\hat{a}_1\hat{a}^\dag_2\hat{a}_2 |0\rangle = 1, \quad \textrm{(using the identities (\ref{identities})).} \] 
which corresponds to the top left hand entry in the matrix (\ref{mat}). In summary, once we know $\bar{U}_{sys}$, we find the expectation value of $\bar{J}_1\bar{K}_2$ by calculating (\ref{mat}) and then reading off the top left corner entry. For the case where $\bar{J}_1 = \bar{I}_1, \bar{K}_2 = \bar{Z}_2$ we obtain the result:
\[ \langle \bar{I}_1\bar{Z}_2 \rangle = 1-2\alpha^2 \xi \, . \]
Depending on whether $\xi$ is closer to zero or one, this term approaches either 
\[ \langle \textbf{0}_1 \textbf{0}_2 | \hat{I}_1\hat{Z}_2 | \textbf{0}_1 \textbf{0}_2 \rangle = 1 \, ,\]
or
\[ \langle \textbf{0}_1 \textbf{0}_2 | \hat{Z}_1\hat{Z}_2 | \textbf{0}_1 \textbf{0}_2 \rangle = 1-2\alpha^2 \, . \]
These two limits correspond to the c-not either not working or working, depending on whether the qubits meet in the gate; that is, on how well their individual amplitudes overlap during the interaction. In this example, according to (\ref{overlap2}), $\xi$ depends on the shift in position $\textbf{x}_1-\textbf{x}_2$ of the pulses.

The key novelty of our formalism is the expression (\ref{mat}), which incorporates the dynamics and spectral features of the qubits into the operator algebra, and the fact that these expressions may be written out explicitly in terms of field operators acting on the global vacuum according to the procedure outlined in section III (i.e. the expression has the form (\ref{heis})). 

\section{Conclusions and future work}
We have developed tools for doing quantum circuit calculations using the particles of a bosonic quantum field for qubits. The algebraic structure of the circuit is defined with respect to a global vacuum state, and expectation values are taken in this state in accordance with the Heisenberg Picture. The usual techniques for calculating evolution on quantum circuits, (matrix representation or Pauli operator algebra) apply here, but now include the method of RMS \cite{ROH07} for dealing with particle mismatch. 

A strong motivation for our work comes from the new area of relativistic quantum information theory (RQI) - concerned with adapting the tools of quantum information to relativistic settings (eg. \cite{TER04,CZA03,CZA08,CZA10,KOK06,HAR95,SCH05,BAR05,RAL09}). Typically, observers in different (especially non-inertial) reference frames can observe inequivalent ground states for fields shared between them \cite{UNR76, ALS03, SCH05}. Calculations for realistic scenarios can rapidly become intractable in the Schr\"odinger Picture due to the highly non-trivial transformations of the ground-state seen by different observers that may be required. In contrast, working in the Heisenberg Picture means the ground-state can be chosen in a suitable inertial frame and all the transformations between frames can be achieved via more straightforward Bogolyubov Transformations of the operators \cite{BIR}.

More generally it is important that formulations of quantum information problems are consistent with current methods for doing quantum field theory on curved spacetimes, particularly the algebraic approach to quantum field theory\cite{WAL06,HOL10}. This approach is based on an algebra of observables constructed from products of field operators, which can take on the interpretation of particle creation and annihilation operators acting on a vacuum state. 
Again, the formalism is expressed in the Heisenberg Picture and expectation values are taken in the appropriate vacuum state. 
The formalism we describe here can be considered as a first step towards developing a generalized qubit algebra that is able to incorporate general transformations in the relativistic setting.

\begin{acknowledgements}
We thank T. Downes and P.P. Rohde for useful discussions. This work was supported by the Australian Research Council.
\end{acknowledgements}

\appendix
\section{Calculation of $\langle\hat{n}_{A_{P}}\rangle$ and $g^{(2)}$}

In this appendix we give the details for calculating the expressions for $\langle\hat{n}_{A_{P}}\rangle$ and $g^{(2)}$ in Eqs.~(\ref{nAp}) and~(\ref{g2}) from Section~\ref{MigdallSourceSection}. We consider the general SPDC evolution given in Eq.~(\ref{Eqn:HeisenbergDC}), using the shorthand notation 
\begin{equation}\label{Eqn:HeisenbergDCApp}
\begin{aligned}
\hat{u}'_{j}&=p \hat{v}_{j}+q \hat{u}^{\dagger}_{j}\\
\hat{v}'_{j}&=p \hat{u}_{j}+q \hat{v}^{\dagger}_{j}
\end{aligned} 
\end{equation}
instead. When calculating $\langle 0|\hat{A}^\dag_{P}\hat{A}_{P}|0\rangle$ and $\langle 0|\hat{A}^\dag_{P}\hat{A}^\dag_{P}\hat{A}_{P}\hat{A}_{P}|0\rangle$ we consider a more general detector model of the form 
\begin{equation}
\hat{d}_{u'_j} \equiv (\mu\hat{\id}-\nu \hat{n}_{u'_j})\hat{n}_{u'_j},
\end{equation} 
which is equivalent to Eq.~(\ref{ffwd}) when $\mu=3/2$ and $\nu=1/2$.

\subsection{Calculation of $\langle 0|\hat{A}^\dag_{P}\hat{A}_{P}|0\rangle$ }

We write the full expression for $\langle 0|\hat{A}^\dag_{P}\hat{A}_{P}|0\rangle$ given $\hat{A}_{P}$ in Eq.~(\ref{ap}): 
\begin{equation}\label{Eqn:nApp}
\langle\sum_{j,k=1}^{N}\prod_{l=0}^{k-1}(1-\hat{d}_{u'_{l}})\hat{v}'^{\dagger}_{k}\hat{d}_{u'_{k}}\hat{d}_{u'_{j}}\hat{v}'_{j}\prod_{i=0}^{j-1}(1-\hat{d}_{u'_{i}})\rangle, 
\end{equation}
where it is understood that all expectation values are taken with respect to the vacuum. The summation  can be broken up into three separate terms:  
\begin{equation}\label{Eqn:SumApp}
\sum_{j,k=1}^{N}=\sum_{j<k=1}^{N}+\sum_{j=k=1}^{N}+\sum_{j>k=1}^{N}.
\end{equation}
Since $\langle  \hat{v}'^{\dagger}_{k}\hat{d}_{u'_{k}}\rangle=0$, only the $\sum_{j=k}$ term in Eq.~(\ref{Eqn:nApp}) survives. Given that
\begin{equation}\label{Eqn:f1f2App}
\begin{aligned}
\langle  \hat{v}'^{\dagger}_{j}\hat{d}_{u'_{j}}\hat{d}_{u'_{j}} \hat{v}'_{j}\rangle&=f_{1}(p,q,\mu,\nu)\\
\langle  (1-\hat{d}_{u'_{i}})^{2} \rangle&=f_{2}(p,q,\mu,\nu),
\end{aligned} 
\end{equation}
where $f_{1}(p,q,\mu,\nu)$ and $f_{2}(p,q,\mu,\nu)$ are both polynomial functions of $p$, $q$, $\mu$ and $\nu$, we find: 
\begin{equation}\label{Eqn:ApDagAp}
\begin{aligned} 
\langle \hat{A}^\dag_{P}(k)\hat{A}_{P}(k)\rangle=f_{1}\sum_{j=1}^{N}\prod_{l=0}^{j-1}f_{2}=f_{1}\left(\frac{f^{N}_{2}-1}{f_{2}-1}\right).
\end{aligned} 
\end{equation}
When we substitute $\mu=3/2$, $\nu=1/2$, $p=1$ and $q=|\chi|$, where $|\chi|\ll 1$ for a SPDC, we obtain Eq.~(\ref{nAp}) to order $|\chi|^{4}$.

\subsection{Calculation of $\langle 0|\hat{A}^\dag_{P}\hat{A}^\dag_{P}\hat{A}_{P}\hat{A}_{P}|0\rangle$ }

We write the full expression for $\langle \hat{A}^\dag_{P}\hat{A}^\dag_{P}\hat{A}_{P}\hat{A}_{P}\rangle$  given $\hat{A}_{P}$ in Eq.~(\ref{ap}) using the fact that $[\hat{v}'^{\dagger}_{i},\hat{d}_{u'_{j}}]=[\hat{v}'_{i},\hat{d}_{u'_{j}}]=0$:
\begin{equation}\label{Eqn:g2App}
\begin{aligned} 
\langle\sum_{{a,c, \atop j,k}=1}^{N}&\hat{v}'^{\dagger}_{k}\hat{d}_{u'_{k}}\hat{v}'^{\dagger}_{a}\hat{d}_{u'_{a}}\hat{d}_{u'_{j}}\hat{v}'_{j}\hat{d}_{u'_{c}}\hat{v}'_{c} \prod_{l=0}^{k-1}(1-\hat{d}_{u'_{l}})\\
&\times \prod_{b=0}^{a-1}(1-\hat{d}_{u'_{b}})\prod_{i=0}^{j-1}(1-\hat{d}_{u'_{i}})\prod_{d=0}^{c-1}(1-\hat{d}_{u'_{d}})\rangle,
\end{aligned} 
\end{equation}
where again it is understood that all expectation values are taken with respect to the vacuum. The summation in this case can be broken up into 75 separate terms, most of which do not contribute:
\begin{itemize}
\item The 24 permutations of summations of the form $\sum_{a>c>j>k}$ do not contribute since $\langle \hat{v}'^{\dagger}_{k}\hat{d}_{u'_{k}}\rangle=0$. 

\item The $12\times3=36$ permutations of summations of the form $\sum_{a=c>j>k}$, $\sum_{a>c=j>k}$ and $\sum_{a>c>j=k}$ do not contribute since $\langle \hat{v}'^{\dagger}_{k}\hat{d}_{u'_{k}} \hat{v}'^{\dagger}_{k}\hat{d}_{u'_{k}}\rangle=0$ and $\langle \hat{v}'^{\dagger}_{k}\hat{d}_{u'_{k}} (1-\hat{d}_{u'_{k}})^{2}\rangle=0$.

\item The $4\times 2=8$ permutations of summations of the form $\sum_{a=c=j>k}$ and $\sum_{a>c=j=k}$ do not contribute since  $\langle \hat{v}'^{\dagger}_{k}\hat{d}_{u'_{k}}  \hat{v}'^{\dagger}_{k}\hat{d}_{u'_{k}} \hat{d}_{u'_{k}} \hat{v}'_{k}\rangle=0$.

\item Four of the 6 permutations of summations of the form $\sum_{a=c>j=k}$ are non-zero and given by
\begin{align} 
\sum_{a=c>j=k}&\langle \hat{v}'^{\dagger}_{a}  \hat{d}_{u'_{a}} \hat{v}'^{\dagger}_{k}  \hat{d}_{u'_{k}}  \hat{d}_{u'_{a}}  \hat{v}'_{a}  \hat{d}_{u'_{k}}  \hat{v}'_{k} \prod_{b=0}^{a-1}(1-\hat{d}_{u'_{b}})^{2} \prod_{l=0}^{k-1}(1-\hat{d}_{u'_{l}})^{2}\rangle\nonumber\\
&=\sum_{a>k} f_{1} f_{4} \prod_{l=0}^{k-1}f_{3}\prod_{b=k+1}^{a-1}f_{2},
\end{align} 
where 
\begin{equation}\label{Eqn:f3f4App}
\begin{aligned}
\langle  (1-\hat{d}_{u'_{k}})^{4} \rangle&=f_{3}(p,q,\mu,\nu)\\
\langle  \hat{v}'^{\dagger}_{k}\hat{d}_{u'_{k}}\hat{d}_{u'_{k}} \hat{v}'_{k}(1-\hat{d}_{u'_{k}})\rangle&=f_{4}(p,q,\mu,\nu)
\end{aligned} 
\end{equation}
are also polynomial functions of $p$, $q$, $\mu$ and $\nu$. 

\item The summation $\sum_{a=c=j=k}$ is non-zero and given by
\begin{align} 
\sum_{a=c=j=k}&\langle \hat{v}'^{\dagger}_{k}  \hat{d}_{u'_{k}} \hat{v}'^{\dagger}_{k}  \hat{d}_{u'_{k}}  \hat{d}_{u'_{k}}  \hat{v}'_{k}  \hat{d}_{u'_{k}}  \hat{v}'_{k} \rangle \langle \prod_{l=0}^{k-1}(1-\hat{d}_{u'_{l}})^{4} \rangle\nonumber\\
&=\sum_{k} f_{5} \prod_{l=0}^{k-1}f_{3},
\end{align} 
where 
\begin{equation}\label{Eqn:f3f4App}
\langle  \hat{v}'^{\dagger}_{k}  \hat{d}_{u'_{k}} \hat{v}'^{\dagger}_{k}  \hat{d}_{u'_{k}}  \hat{d}_{u'_{k}}  \hat{v}'_{k}  \hat{d}_{u'_{k}}  \hat{v}'_{k}  \rangle=f_{5}(p,q,\mu,\nu)
\end{equation}
is a polynomial functions of $p$, $q$, $\mu$ and $\nu$. 
\end{itemize}

The expression for $\langle \hat{A}^\dag_{P}\hat{A}^\dag_{P}\hat{A}_{P}\hat{A}_{P}\rangle$ therefore reduces to:

\begin{align} 
 & 4\sum_{a>k} f_{1} f_{4} \prod_{l=0}^{k-1}f_{3}\prod_{b=k+1}^{a-1}f_{2}+\sum_{k=1}^{N} f_{5} \prod_{l=0}^{k-1}f_{3}\label{Eqn:g2FinalApp}\\
&=4 f_{1} f_{4}\left(\frac{f_{3}-f_{2}+f_{3}^{N}(f_{2}-1)-f_{2}^{N}(f_{3}-1)}{(f_{2}-1)(f_{3}-1)(f_{3}-f_{2})}\right)+f_{5}\left(\frac{f_{3}^{N}-1}{f_{3}-1}\right).\nonumber
\end{align} 
When we substitute $\mu=3/2$, $\nu=1/2$, $p=1$ and $q=|\chi|$, where $|\chi|\ll 1$ for a SPDC, we obtain Eq.~(\ref{g2}) to order $|\chi|^{4}$.

\bibliographystyle{apsrev4-1}

\end{document}